\documentclass[12pt,preprint]{aastex}

\slugcomment{}
\usepackage{bm}
\usepackage{graphicx}

\shorttitle{Y.Uchida et al.}
\shortauthors{Faraday Rotation Measure Predicted from MHD Model I}

\begin{document}

\title{Distribution of Faraday Rotation Measure \\
in Jets from Active Galactic Nuclei \\
I. Prediction from our Sweeping Magnetic Twist Model}

\author{Yutaka Uchida\altaffilmark{1}, Hiromitsu
Kigure\altaffilmark{1,2}, 
Shigenobu Hirose\altaffilmark{1}, 
Masanori Nakamura\altaffilmark{1,3}, and Robert Cameron\altaffilmark{1}} 
\altaffiltext{1}{Department of Physics, Science University of Tokyo, 1-3 Kagurazaka, Shinjuku-ku, Tokyo 162-8601, Japan}
\altaffiltext{2}{Kwasan and Hida Observatories, Kyoto University,
Yamashina, Kyoto 607-8471, Japan;kigure@kwasan.kyoto-u.ac.jp}
\altaffiltext{3}{Jet Propulsion Laboratory, California Institute of Technology, 4800 Oak Grove Drive, Pasadena, CA 91109, USA}

\begin{abstract}
  Using the numerical data of MHD simulation for AGN jets 
  based on our ``sweeping magnetic twist model'', we calculated the 
  Faraday rotation measure (FRM) and the Stokes parameters to compare 
  with observations. 
  We propose that the FRM distribution can be used to discuss the 
  3-dimensional structure of magnetic field around jets, together with the 
  projected magnetic field derived from the Stokes parameters.  
  In the present paper, we supposed the basic straight part of AGN jet, 
  and used the data of axisymmetric simulation. The FRM distribution 
  we derived has a general tendency to have gradient across the jet axis,
  which is due to the toroidal component of the helical magnetic field 
  generated by the rotation of the accretion disk. 
  This kind of gradient in the FRM 
  distribution is actually observed in some AGN jets (e.g. Asada et
  al. 2002), which suggests helical magnetic field around the jets and
  thus supports our MHD model. Following this success, we are now
  extending our numerical observation to the wiggled part of the jets
  using the data of 3-dimensional simulation based on our model in the
  following 
  paper. 
\end{abstract}


\section{Introduction}
To explain the formation of active galactic nucleus (AGN) jets and 
other astrophysical jets, various models have been proposed. Among them, 
magnetohydorodynamic (MHD) model is one of the most promising models, 
since it can explain both the acceleration and the collimation of the jets. 
Lovelace (1976) and Blandford (1976) first proposed the magnetically
driven jet from accretion disks, and 
Blandford \& Payne (1982) discussed magneto-centrifugally driven
outflow from a Keplerian disk in steady, axisymmetric and self-similar 
situation.
Uchida \& Shibata (1985) performed a time-dependent, two-dimensional
axisymmetric simulation in the case of star-forming outflows. They
pointed out that large amplitude torsional Alfv\'en waves (TAW's)
generated by the interaction between the accretion disk and a large
scale magnetic field play an important role (detail is described in
section \ref{sec:review-model}). In this paper, we refer this model as
"sweeping magnetic twist model". Uchida \& Shibata (1986) extended the
treatment to the case of AGN jets. After this work, many authors
have performed time-dependent, two-dimensional axisymmetric simulations 
(e.g. Stone \& Norman 1994, Ustyugova et al. 1995, Matsumoto et
al. 1996, Ouyed \& Pudritz 1997, Kudoh, Matsumoto, \& Shibata 1998). 
Acceleration mechanisms in MHD model were studied in detail by 
1.5-dimensional MHD equations (Kudoh \& Shibata 1997a, 1997b).

Using the numerical data of MHD model,
observational quantities such as the Faraday rotation measure (FRM) or 
the Stokes parameters have been derived 
to compare with observations of AGN jets: 
Laing (1981) computed the
total intensity, the linear polarization, and the projected magnetic
field distributions, assuming some simple magnetic field configurations 
and high energy particle distributions in the cylindrical jet. Clarke, 
Norman, \& Burns (1989) performed two dimensional MHD
simulations in which a supersonic jet with a dynamically passive helical
magnetic field was computed, and derived distributions of the total 
intensity, the projected electric field, and the linear polarization. 
Hardee \& Rosen (1999) calculated the total intensity and the projected
magnetic field distributions, using 3-dimensional MHD simulations of
strongly magnetized conical jets. Hardee \& Rosen (2002)
calculated the FRM distribution and discussed
that the radio source 3C465 in Abell cluster A2634 (Eilek \& Owen 2002)
suggests helical twisting of the flow.

The FRM is given by the integral of $n_e B_\parallel$ along 
the line-of-sight between the emitter and the observer (where
$B_\parallel$ is the line-of-sight component of the magnetic field, and
$n_e$ is the electron density there). It is, in principle, not possible
to specify which part on the line-of-sight the contribution comes
from. However, in recent high-resolution radio observations (e.g. Eilek \& Owen
2002, Asada et al. 2002), the FRM distribution seems to have good 
correlation with the configuration of the jet; 
this suggests that the FRM
variation is due to the magnetized thermal plasma surrounding
the emitting part of the jet. In fact, sharp FRM gradients seen in
3C273 can not be produced by a foreground Faraday screen
(Taylor 1998, Asada et al. 2002).
If this is the case, 
we can get a new information, that is, the line-of-sight component of 
the magnetic field, and thus can predict 
the 3-dimensional configuration of the
magnetic field around the jet, together with the projected magnetic field.

In this paper, we calculate the FRM, projected magnetic field, and 
total intensity from the numerical data of MHD simulation 
based on our ``sweeping magnetic twist model'', and discuss these model 
counterparts comparing with some observations. Here we consider the 
straight part of the jet, and thus use the data of axisymmetric simulation. 
In section 2, we review the physics of 
our "sweeping magnetic twist model". We introduce the 
method to calculate model counterparts of observational quantities 
in section 3, and show the results in section 4. 
Comparisons of model counterparts with some observations 
are discussed in section 5.

\section{Brief Review of Our ``Sweeping Magnetic Twist Model''}\label{sec:review-model}
In this section, we briefly review the results of 2.5-dimensional 
MHD simulations based on our ``sweeping magnetic twist model'' 
to discuss the magnetic field around the 
straight part of jets. In the following paper, 
we will extend our treatment to the wiggled part of jets, which we 
have given an interpretation using a 3-dimensional MHD simulation 
based on our model (Nakamura, Uchida, \& Hirose 2001).

In the original MHD model (Uchida \& Shibata 1985) for 
bipolar outflows in star-forming regions, they considered a gravitational 
contraction of magnetized gas to form a star (plus an accretion disk).
They attributed 
the large scale magnetic field to the weak field in the Galactic arms. 
It is strengthened in the process of gravitational 
contraction of the interstellar gas to the star-forming core, 
and plays a critical role. The toroidal field is continuously produced 
from the poloidal field by the rotation of the accretion disk. 
This causes magnetic braking to the disk material,
and the material which loses angular momentum falls 
gradually toward the central gravitator, 
and releases the gravitational energy. 
A part of the released gravitational energy is supplied to the jets 
along the magnetic field. 
The produced toroidal magnetic field propagates into two
directions along the bunched large scale 
magnetic field as large amplitude TAW's. These TAW's 
serve to collimate the large scale poloidal field
into the shape of a slender jet by dynamically pinching
it in the propagation (``sweeping pinch effect''). This process, 
verified in the simulation, was proposed by Uchida \& Shibata (1985) as
a generic magnetic effect operating in the formation of astrophysical
jets utilizing gravitational energy.

The mechanism was applied to the case of AGN jets (Uchida \& Shibata
1986) by supposing that a large scale intergalactic magnetic field plays
the role in the case of the formation of a protogalaxy and a giant black
hole at its core. They argued that the same process as in the star
formation case is applicable to the AGN jet cases with more or less
similar set up (having accretion disk around the central gravitator
etc.), due to the similarity of the basic equation system.
One of the possible differences between AGN jets and the 
star formation jets may be the relativistic effects.
The effect of general relativity
will be appreciable very close to the central giant black 
hole comparable to the 
Schwarzschild radius (Koide, Shibata, \& Kudoh 1998). 
There are regions in which 
the special relativity should be taken into account
when the Alfv\'en velocity estimated in the classical definition 
is close to or exceed the velocity of light. Here in this paper,
we concentrate ourselves on the essential physical process 
in the production and collimation of the jet in the non-relativistic
range.

The problem was treated with the non-linear system of
MHD equations in a time dependent way 
for the first time when they proposed this model in 1985. 
The numerical approach was so-called 
axial 2.5-dimensional approximation, where the quantities are axisymmetric, 
but the azimuthal components of vectors are included to allow them play 
very essential roles such as centrifugal effect or pinch effect. 
Thus the authors were able to deal 
with the physical driving and collimating mechanism they proposed to
be in operation for astrophysical jets.

\begin{figure}
\figurenum{1}
\epsscale{0.4}
\plotone{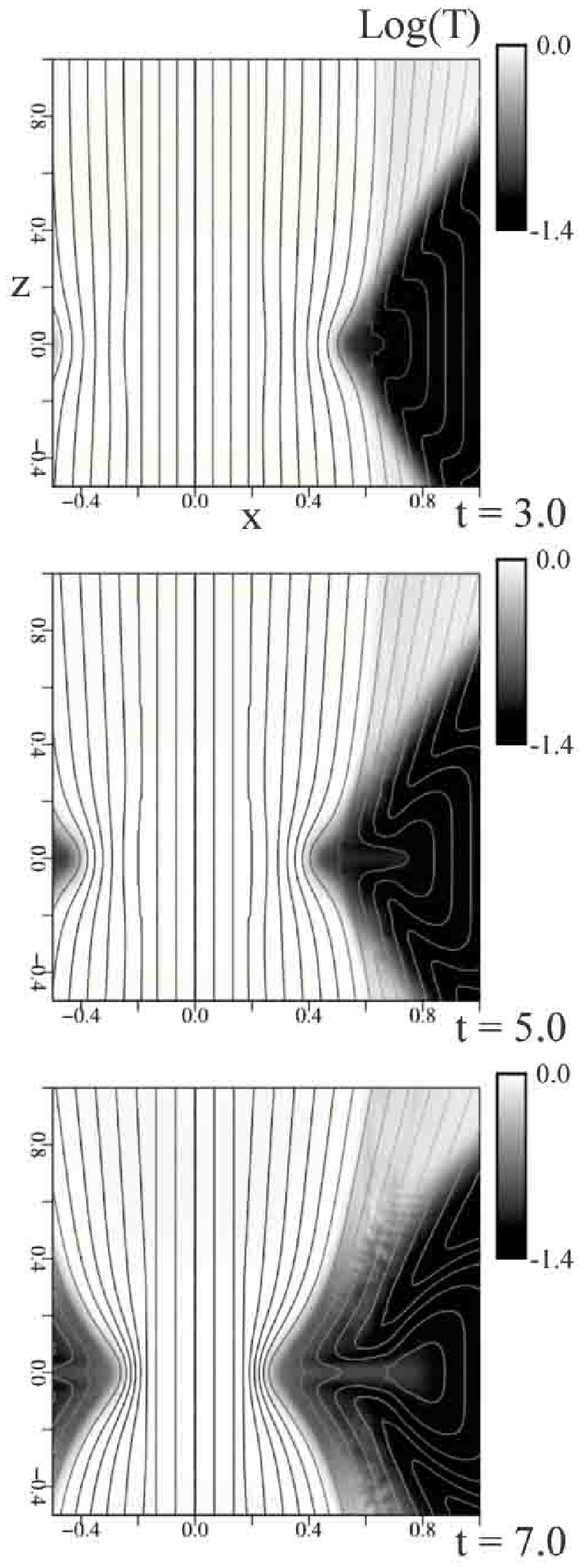}
\caption{Time development in the $r-z$ plane in our 
2.5-dimensional MHD simulation near the central core. 
The distribution of logarithmic 
temperature is shown in gray-scale. Curves represent the 
poloidal magnetic field. 
\label{FIG01}} 
\end{figure}

\begin{figure}
\figurenum{2}
\epsscale{0.9}
\plotone{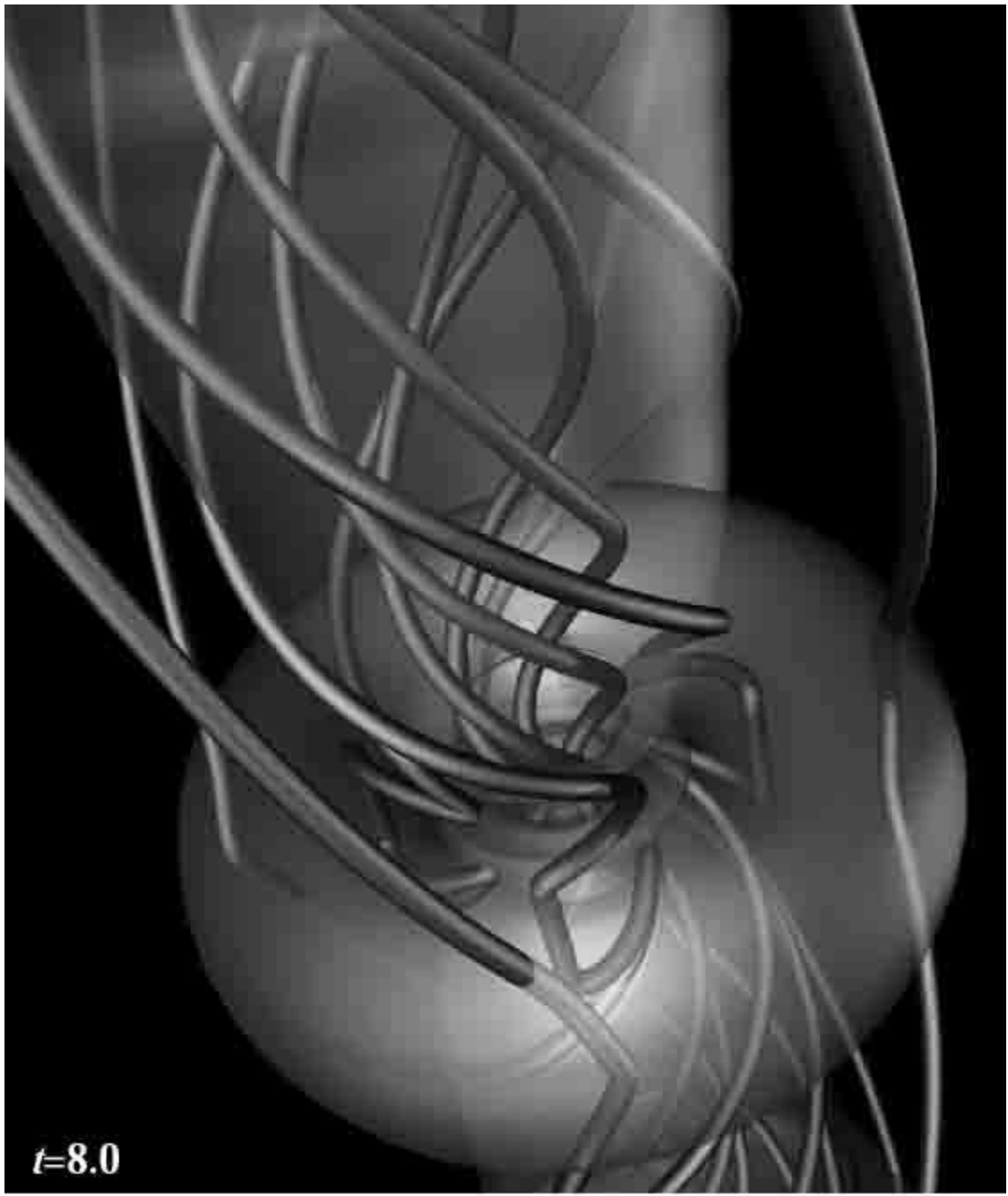}
\caption{3-dimensional presentation of the magnetic field with the
 accretion torus and jets (Meier, Koide, \& Uchida 2001, Front page of
 Science). 
\label{FIG02}} 
\end{figure}

Figure \ref{FIG01} shows the time development in the 2.5-dimensional MHD
simulation based on our model. The rotating gas pulls the
magnetic field gradually inward, which twists up the magnetic field
because the rotational velocity is faster as close to the center (Figure
\ref{FIG02}). This continuously supplies large amplitude TAW's (Poynting
flux) along the external magnetic field, which pinch the poloidal
magnetic field into the shape of a slender jet as discussed in the 
above. The gas in the surface of the torus is swirled 
out into two directions along the axis, both by the magnetic pressure
gradient and the centrifugal effect. Thus the propagation of the
torsional Alfv\'en wave accelerates the gas in the surface of the disk
into the spinning jets. 

It is noted that the accretion toward the
central gravitator takes place in the form of avalanches from upper
and lower surfaces of the geometrically thick torus (Matsumoto et
al. 1996), because the transfer of angular momentum to the external 
magnetic field is most efficient there.
The magnetic fields of opposite polarity, brought with the accretion 
flows avalanching on the surfaces of the disk,
make reconnection at the innermost edge of the 
disk in the equatorial plane (Figure \ref{FIG01}). This process will 
contribute to the supply of 
the ``seed high energy particles'' into the jet. 
Such particles will be re-accelerated through 
the Fermi-I acceleration process 
when two TAW fronts trapping the particles in between them
approach to each other, for example, as the foregoing one is decelerated  
due to an encounter with high density gas blob remaining in the
collapse (Uchida et al. 1999).

\section{Method of Calculation of Model Counterparts}
Using the numerical data of the 2.5-dimensional simulation explained 
in the previous section, 
we computed the distributions of the FRM, the projected 
magnetic field, and the total intensity with some viewing angles.  

We computed the FRM distribution by integrating $n_e
B_\parallel$ along the line-of-sight (Hardee \& Rosen 2002). 
To calculate the Stokes parameters, we assume the following:
(1)radiation process is synchrotron radiation, (2)synchrotron 
self absorption is negligible, (3)the spectral index, $\alpha$, 
is equal to unity, and (4)the projected magnetic field is 
perpendicular to the projected electric field.
The emissivity of the synchrotron radiation is given by 
$\epsilon = p |B \sin \psi|^{\alpha + 1}$,
where $B$ is the local magnetic field strength, $\psi$ is the
angle between the local magnetic field and the line-of-sight, and $p$ is
the gas pressure. In our simulation the relativistic particles are not
explicitly tracked, therefore we assume that the energy and number
densities of the relativistic particles are proportional to the energy
and number densities of the thermal fluid (Clarke et al. 1989, Hardee \&
Rosen 1999, 2002). The total intensity is then given by the integration
of the emissivity along the line-of-sight as 
$I = \int \epsilon ds $. Other Stokes parameters 
are given by $Q = \int \epsilon \cos 2 \chi^{'} ds$ and 
$U = \int \epsilon \sin 2 \chi^{'} ds$,
where the local polarization angle 
$\chi^{'}$ is determined by the direction of the local magnetic field 
and the direction of the line-of-sight. Using these $U$ and $Q$, the
polarization angle $\chi$ is given by 
$\chi = (1/2)\tan^{-1}(U/Q)$. Finally the projected magnetic field is 
determined from the polarization angle
$\chi$ and the polarization intensity $\sqrt{Q^2 + U^2}$.

\begin{figure}
\figurenum{3}
\epsscale{1.0}
\plottwo{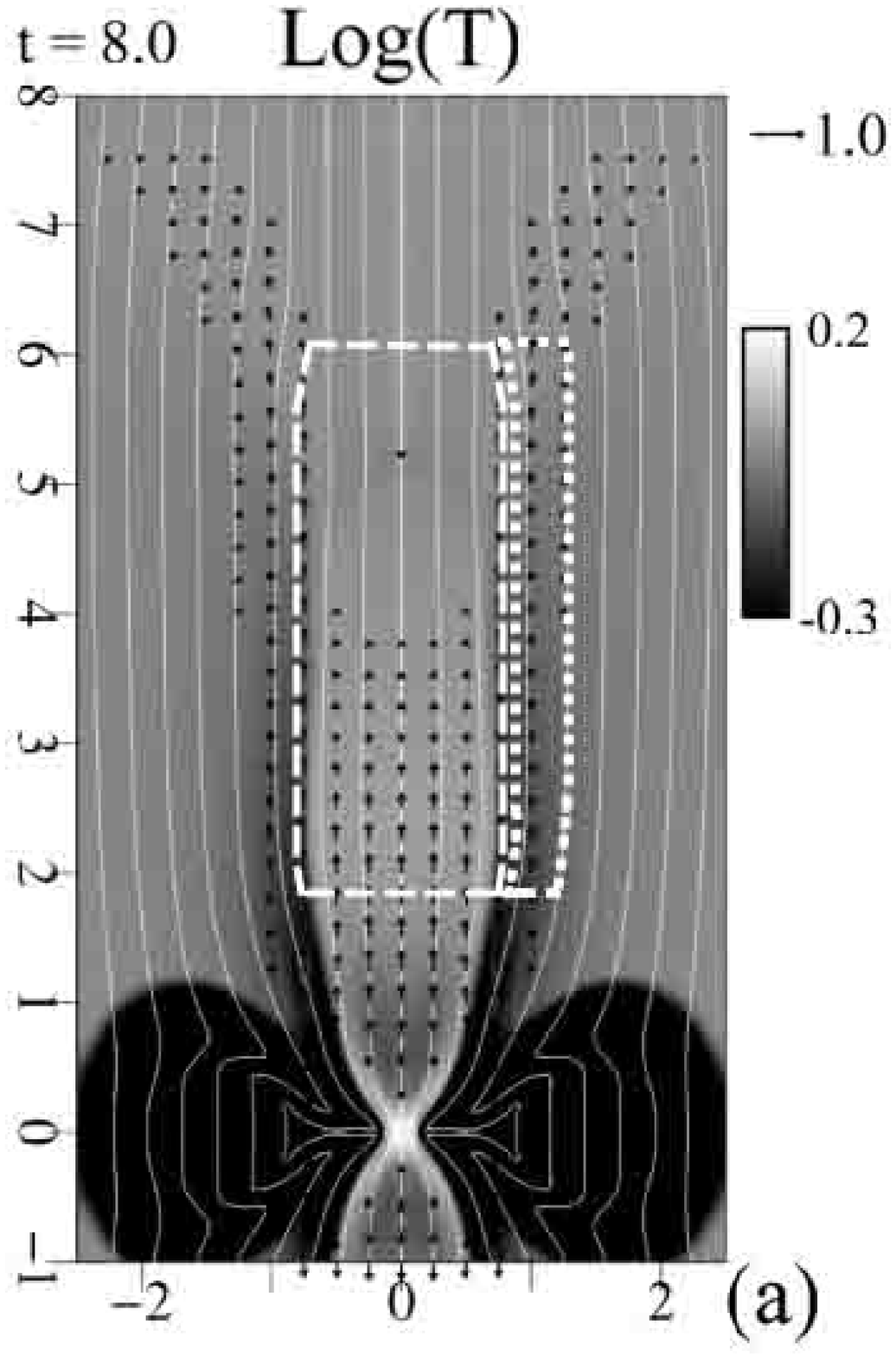}{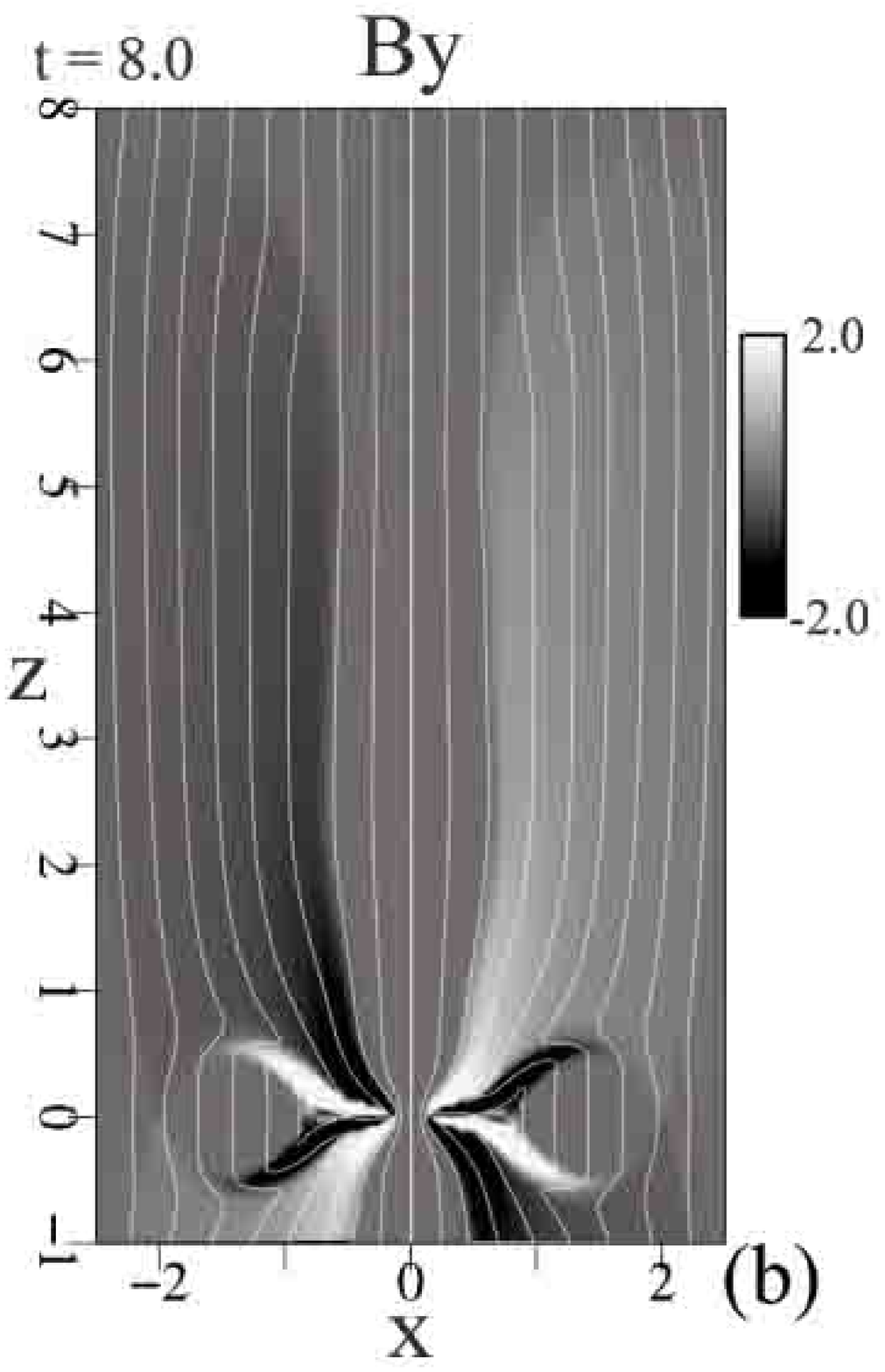}
\caption{(a) The distribution of logarithmic temperature
 (gray-scale, dark is lower temperature) together with the
 poloidal field (white curves) and poloidal velocity field (arrows) in
 the $x-z$ plane. The white box with dotted lines, and that with dashed
 lines are assumed to be the Faraday rotation screen, and the emission
 region, respectively. (b) The poloidal magnetic field and the toroidal
 magnetic field ($B_y$) (gray-scale, white is plus and dark is minus).
\label{FIG03}}
\end{figure}

Here we separate the Faraday rotation screen and the emitting region, 
and we performed the integrations only in the 
emitting region for the Stokes parameters, 
and only in the Faraday rotation screen for the FRM.
We assumed this separation on the basis of the fact that 
linear dependence of the observed
polarization angle on wavelength-squared holds in 
some observations (Perley, Bridle, \& Willis 1984, Feretti et al. 1999,
Asada et al. 2002); this would not be the case if the Faraday rotation
is caused in the emitting region (Burn 1966). Figure \ref{FIG03} shows
the emitting region (the region in the box of dashed lines) and the
Faraday rotation screen (the region in the box of dotted lines) 
assumed in our calculations. 
The cylindrical shell outside the emitting
region is assumed to play the main role of the Faraday rotation screen,
because the temperature is lower and the toroidal field is stronger
there (Figure \ref{FIG03}) 
compared with those in the tenuous clouds in the intergalactic
space. 

We consider two types of the emitting regions; one is the layer type 
(type L: high energy particles exist only in the skin part of the 
dashed box in Figure \ref{FIG03}(a)), and the 
other is the column type (type C: high energy particles fill the whole 
region of the dashed box). 
The former corresponds to the idea that the
high energy particles are injected into the inner skin part of the 
jet due to the magnetic
reconnection at the inner edge of the accretion disk as described in
the previous section. The latter may happen if the high energy particles 
come from the pair plasma creation in the black hole magnetospheres.

\section{Results of Numerical Observations}
\subsection{Faraday Rotation Measure}
The model counterparts of the FRM distribution with different viewing
angles are shown in Figure \ref{FIG04}. When we see the jet from the
direction perpendicular to its axis ($\theta=90^\circ$, $\theta$ is the
angle between the jet axis and the line-of-sight), we see only the
toroidal component of the helical field and thus the value of the FRM is
almost {\it antisymmetric} with respect to the axis (it is not perfectly 
antisymmetric because the radial component of the magnetic field is not
equal to zero). The distribution of the FRM is distorted from
antisymmetry as the viewing angle varies, but it {\it always shows
gradient across the jet axis} (Figure \ref{FIG04}-2); this gradient
across the jet axis can be interpreted as the sum of the antisymmetric
part due to the toroidal component of the helical field and a base-value
due to the longitudinal component. When $|\theta-90^\circ|$ is larger
than the pitch angle of the helical field, this longitudinal component
dominates (Figure \ref{FIG04}-1).

\subsection{Projected Magnetic Field and Total Intensity}
Figure \ref{FIG05} shows the distributions of the projected magnetic
field and the total intensity with different viewing angles in the case
of the type L emitting region. When $\theta$ is equal to $90^\circ$, 
the total intensity is nearly constant, but has an edge-brightening; 
this is because the emissivity becomes smaller away from the axis and 
the integration depth of the emitter has a maximum at edges. 
In other cases, the 
distribution of the total intensity is asymmetric, 
since the emissivity changes as 
$|\sin \left( \theta - \zeta  \right)|^{\alpha+1}$ at the left edge and
$|\sin \left( \theta + \zeta  \right)|^{\alpha+1}$ at the right edge, where 
$\zeta$ is the pitch angle of the clockwise (seen from the jet origin) 
helical field.\footnote{Note that this is opposite sense to that shown in 
Figure \ref{FIG02}.}
Therefore, for example, it becomes dark at the left edge 
when $\theta$ is nearly equal to $\zeta$.
As for the projected magnetic field, it is perpendicular to the jet
axis in the almost entire region, which does not depend on the viewing angle. 
This is because the the toroidal magnetic field is dominant in the emitting 
region. 

Figure \ref{FIG06} shows the distributions in the case of the type C
emitting region. In this case, the intensity has the peak on 
the axis when $\theta=90^\circ$, because the integration depth has 
a maximum at the center of the jet. When the viewing angle is not so
small, the projected magnetic field is parallel to the jet axis, which 
corresponds to the poloidal magnetic field in the emitting region. 
When the viewing angle is small (e.g. $\theta=15^\circ$), the 
distributions of the total intensity and the projected magnetic field
are almost same as those in the type L. This is because the fraction of
the magnetic field perpendicular to the line-of-sight, which contributes
to the synchrotron radiation, is small in both cases.

\section{Summary and Discussion}
We calculated the FRM and the Stokes parameters using numerical 
data of 2.5-dimensional MHD simulation based on our ``sweeping
magnetic twist model''. The distribution of the FRM always 
has a gradient across the jet axis, which is
caused by the toroidal magnetic field generated by the disk rotation. 
In calculating the Stokes parameters, we assumed two types of emitting 
regions, the column type and the layer type. In the former, the projected 
magnetic field tends to be parallel to the jet axis and the total intensity 
has the peak at the jet center. On the other hand, in the latter, 
the projected magnetic field is perpendicular to the axis; the 
total intensity has ``edge-brightening'', which is observed in
Centaurus A (Clarke, Burns, \& Feigelson 1986), M87 (Biretta, Owen, \&
Hardee 1983, Junor, Biretta, \& Livio 1999). In the following, 
we discuss the 3-dimensional magnetic field structure around the 
observed jets using these results of our numerical observations, 
especially focusing on the FRM distribution. 

Figure \ref{FIG07} shows a recent observational result for 3C273 jet
obtained by Asada et al. (2002) by using the VLBA Archive Data. In this
case, the FRM distribution has a systematic gradient in the direction 
perpendicular to the jet axis, which can be interpreted as 
the sum of an antisymmetric
distribution and a base value (a constant over the source). 
It is not likely that the foreground magnetized cloud at
a large distance from the jet has a very sharp variation
of either magnetic field or the density, just along the projected 
very thin jet (Taylor 1998, Asada et al. 2002). One likely 
interpretation may be that the antisymmetric contribution is due to the
toroidal component of the helical magnetic field as stated in the above. 
The base-value may come either from the foreground 
large scale magnetized clouds at
large distances, or from the longitudinal 
component of the helical magnetic field. 
The latest observation of the 3C273 jet shows that the systematic FRM
gradient persists along a more significant length of the jet 
(Asada, private communication), which would be a strong evidence 
for the above idea. 

\begin{figure}
\figurenum{4}
\epsscale{1.0}
\begin{center}
\rotatebox{90}{\plotone{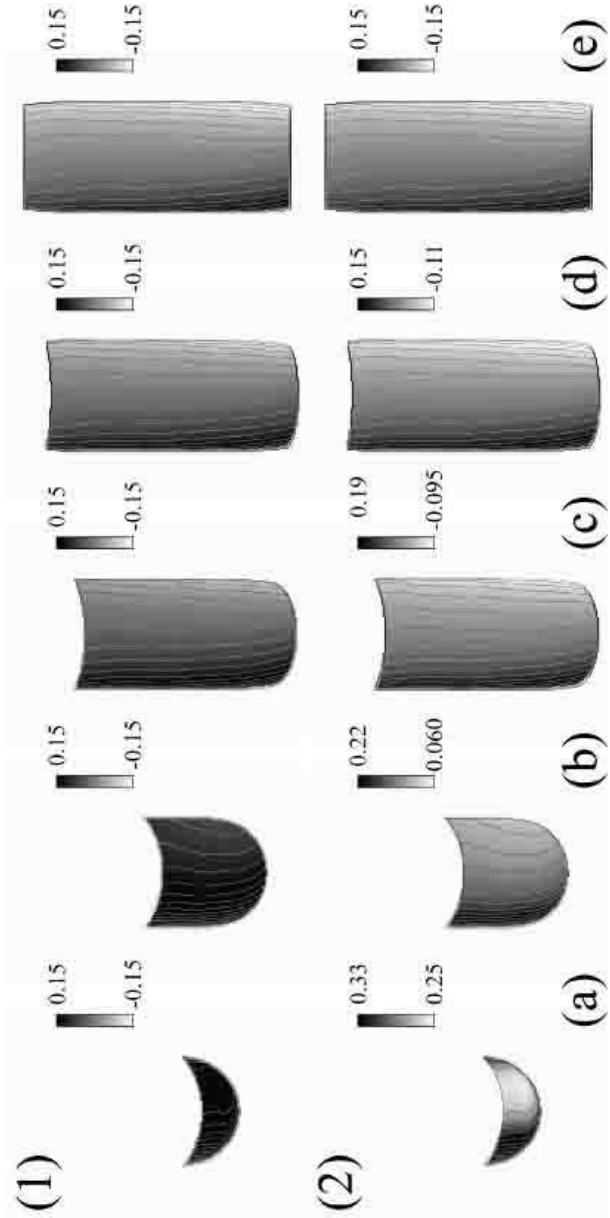}}
\caption{Calculated model counterparts for the FRM distribution, when
 seen at (a) $15^\circ$, (b) $30^\circ$, (c) $60^\circ$, (d) $75^\circ$,
 and (e) $90^\circ$, from the axis (ahead of the jet). 
 Only those corresponding to the upper part of the jet ($z>0$) 
 are shown. The direction of the jet coincides with that of the
 longitudinal  magnetic field. The gray-scale range of the FRM
 distribution is fixed in the 1st row,  while it is determined in each
 viewing angle in the 2nd row. The accretion disk is lower side in these
 pictures.\label {FIG04}} 
\end{center}
\end{figure}

\begin{figure}
\figurenum{5}
\epsscale{1.0}
\begin{center}
\rotatebox{90}{\plotone{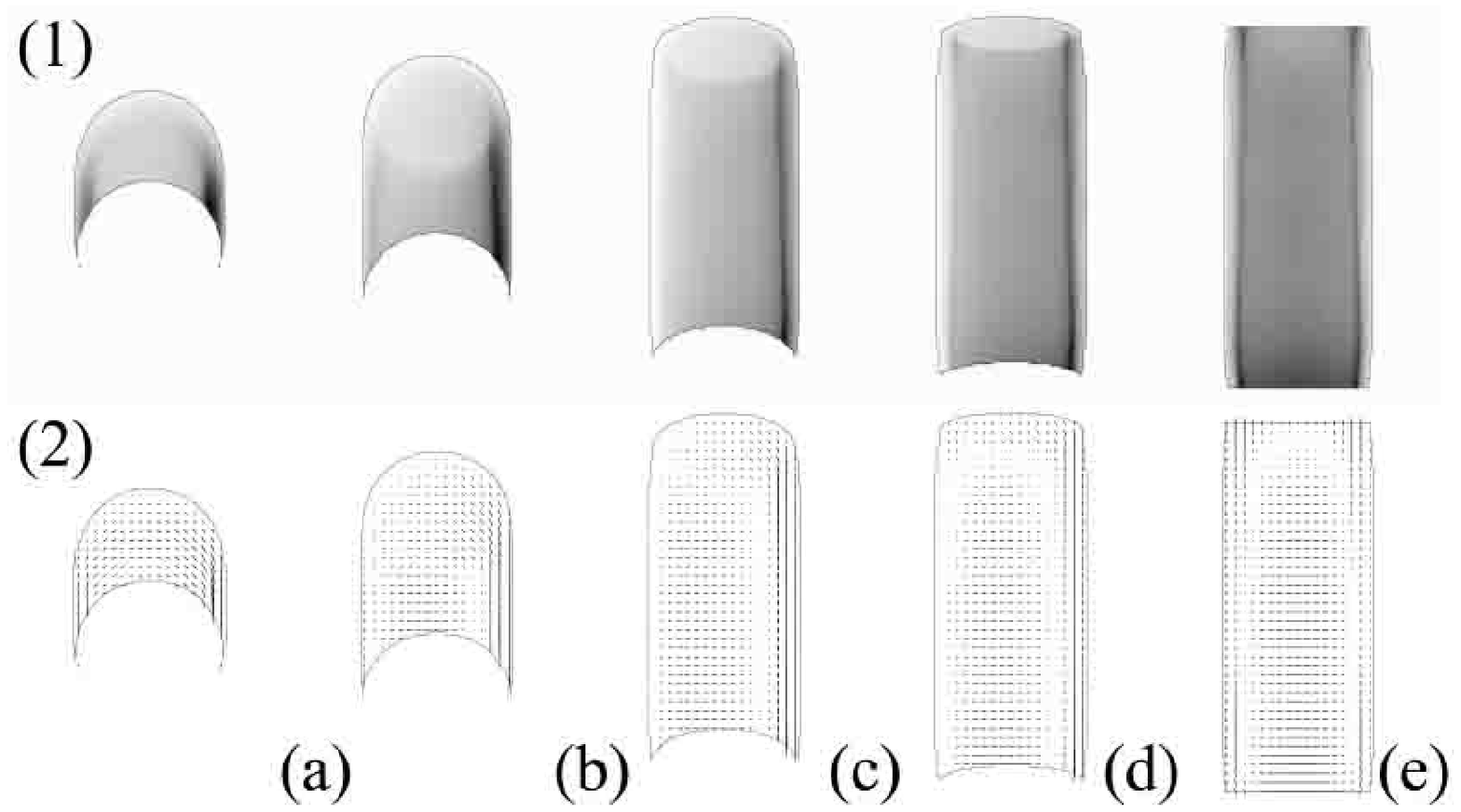}}
\caption{Calculated model counterparts for the total intensity and
 the projected magnetic field in the case of the type L emitting region, 
 when seen at (a) $15^\circ$, (b) $30^\circ$, (c) $60^\circ$, (d)
 $75^\circ$, and (e) $90^\circ$, from the axis. The black in the gray
 scale indicates the maximum and the white indicates the minimum.
\label{FIG05}} 
\end{center}
\end{figure}

\begin{figure}
\figurenum{6}
\epsscale{1.0}
\begin{center}
\rotatebox{90}{\plotone{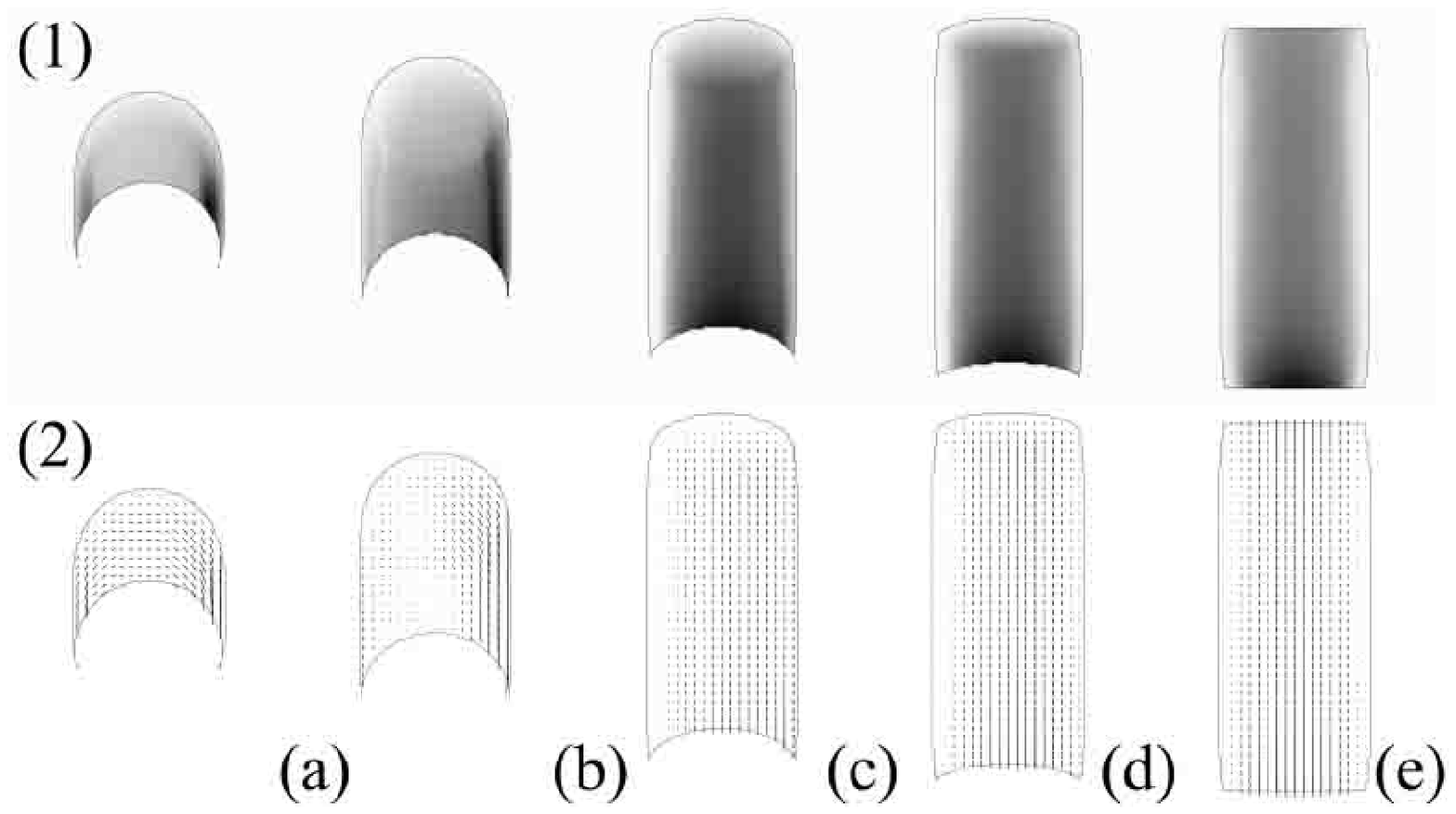}}
\caption{Calculated model counterparts for the total intensity and
 the projected magnetic field in the case of the type C emitting region, 
 when seen at (a) $15^\circ$, (b) $30^\circ$, (c) $60^\circ$, (d)
 $75^\circ$, and (e) $90^\circ$, from the axis. The black in the gray
 scale indicates the maximum and the white indicates the minimum.
\label{FIG06}}
\end{center}
\end{figure}

\begin{figure}
\figurenum{7}
\epsscale{0.6}
\plotone{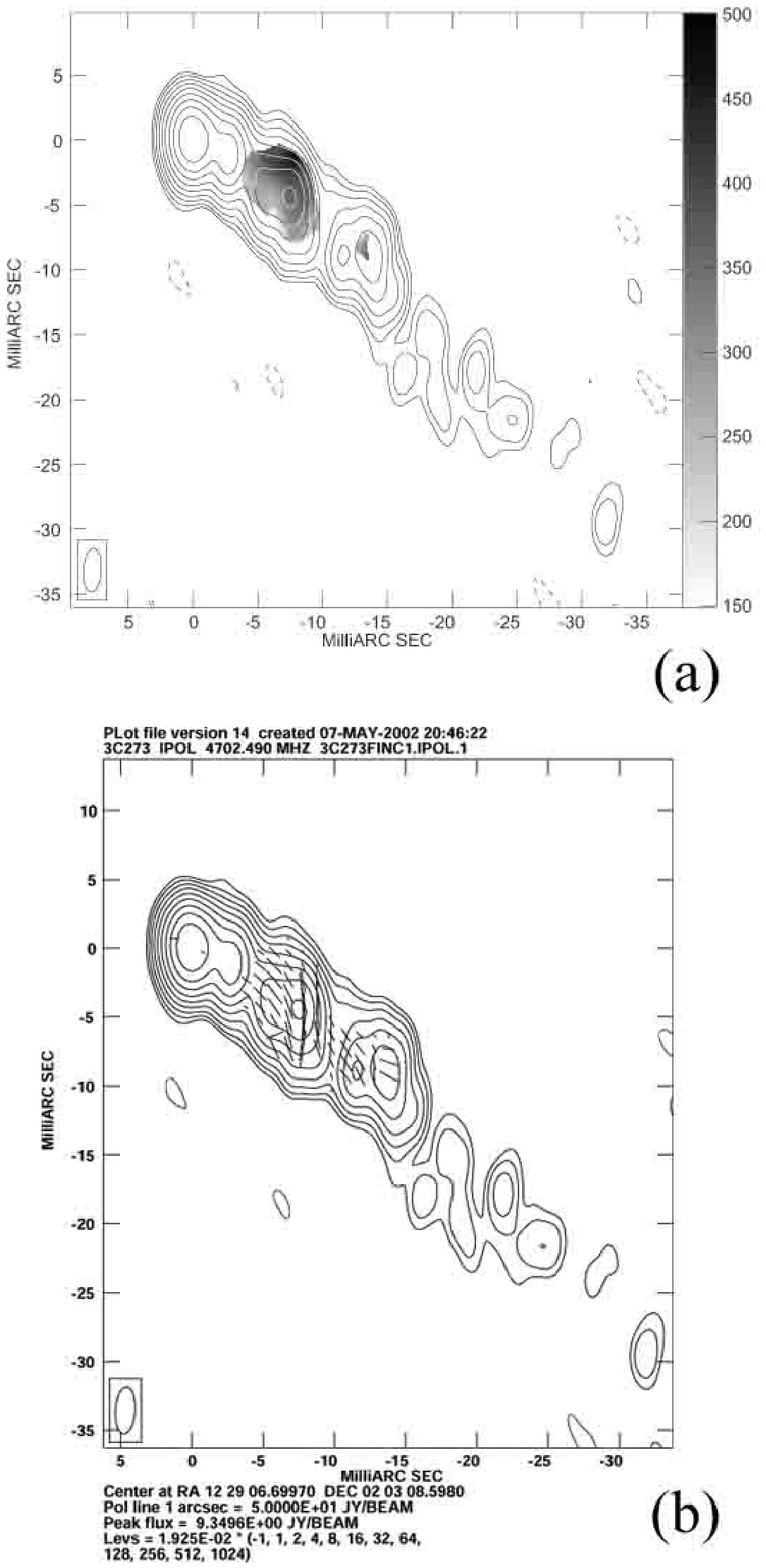}
\caption{The distributions of (a) the Faraday rotation measure (this was
 modified from color picture to monochrome by courtesy of the authors),
 and (b) the direction of projected magnetic field, in the 3C273 jet
 (Asada et al. 2002).\label{FIG07}} 
\end{figure}

The projected magnetic field in 3C273 
is generally tilted somewhat from the axis of the jet, and the 
tilt angle becomes large at the blob called ``anomaly''. 
Asada et al. (2002) attributed this to the shocks created on encountering
non-uniformity. Another possibility is that the jet {\it is bent} at that
point so that the jet axis has less angle from the line-of-sight; the 
change of the FRM value can be explained with our model counterparts of 
different viewing angles (see the change, for example, from (c) to (b) 
in Figure \ref{FIG04}(1)).

\begin{figure}
\figurenum{8}
\epsscale{1.0}
\plotone{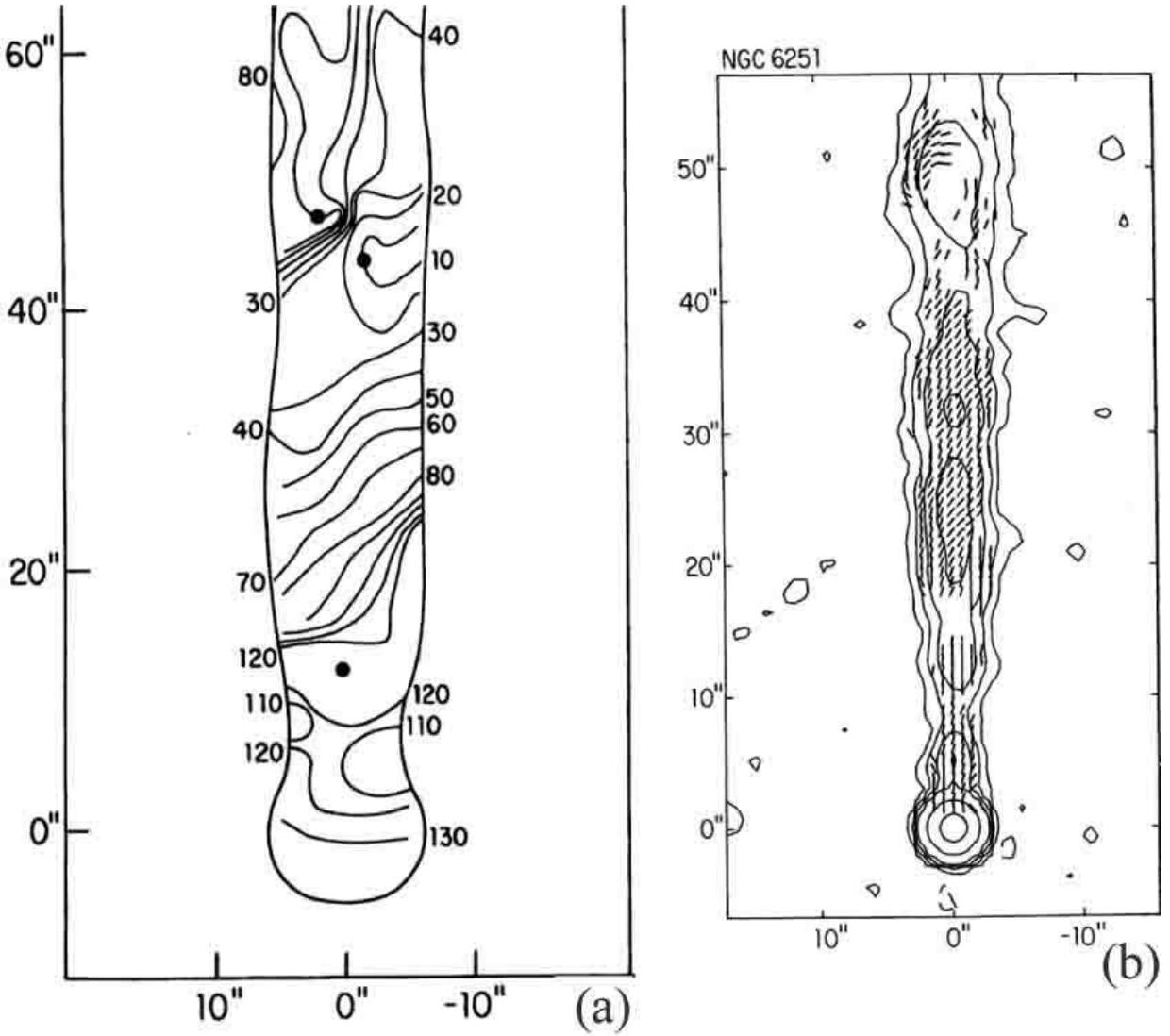}
\caption{The distributions of (a) the Faraday rotation measure, 
and (b) the direction of projected magnetic field,  
in NGC 6251 jet (Perley et al. 1984). 
\label{FIG08}} 
\end{figure}

The observation of the jet from the AGN core of NGC6251 
(Perley et al. 1984) is shown in Figure \ref{FIG08}. This
may be a bit weaker evidence, but still considerably positive evidence
for the systematic twist in the magnetic field. In this case, the
contour lines of the FRM are nearly parallel in the central region up to
40\arcsec; therefore the distribution of the FRM has a systematic
antisymmetry with a gradient in the direction perpendicular to the jet
axis, if we subtract the possible contribution from the spherically
condensed gas cloud probably associated with the central part of the
host galaxy. The projected magnetic field is aligned and this would be
consistent with the propagation of the torsional Alfv\'en waves.

The systematic and gradual change of the FRM value across the jet axis
as seen in the above examples might suggest the existence of 
helical magnetic field. In the MHD models,
it is naturally explained as the results of the interaction between 
the large-scale magnetic field and the disk rotation. 
Non-magnetic models may not be able to explain the FRM
gradient. Although hydrodynamic models can include magnetic field as a
passive ingredient, the magnetic field in such a case is expected to be
carried around passively by the motions of non-magnetic gas
dynamics. Such passively distorted magnetic field will produce the
Faraday depolarization, rather than showing clear systematic
FRM distribution. 

The FRM distribution, when the Faraday rotation occurs 
in the external medium around the jet, can be useful to determine 
the 3-dimensional magnetic field structure around the jet, together 
with the projected magnetic field. We have 
demonstrated that the characteristic 
distribution of the FRM (systematic gradient perpendicular to the jet
axis) in the straight part of some AGN jets can be explained 
by our ``sweeping magnetic twist model''. 
On the basis of this success, we are extending our numerical observation 
to the wiggled 
part of jets, which can be explained as a structural helix produced 
by MHD kink instability in the regime of our model (Nakamura et al. 2001). 
The results will be reported in the following paper.

\acknowledgments
We acknowledge Mr. K. Asada for providing their precious results of the
FRM distribution, and Prof. M. Inoue for discussion. We hope that more
observations of the FRM distribution in AGN jets will be performed in
the near future, since they are very important in determining the
correct theoretical model. Numerical computations were carried out on
VPP5000 at the Astronomical Data Analysis Center of the National
Astronomical Observatory, Japan, which is an interuniversity research
institute of astronomy operated by the Ministry of Education, Culture,
Sports, Science, and Technology.

 

\clearpage

\clearpage

\clearpage


\end{document}